\documentclass[showpacs,twocolumn,prl,aps]{revtex4}
\usepackage{graphicx}

\begin{document}

\title[Short title for running header]{An efficient Monte Carlo algorithm for the evaluation of Renyi entanglement entropy of a general quantum dimer model at the R-K point}
\author{Jiquan Pei, Qiang Han, Haijun Liao and  Tao Li}
\affiliation{ Department of Physics, Renmin University of China,
Beijing 100872, P.R.China}
\date{\today}

\begin{abstract}
A highly efficient and simple to implement Monte Carlo algorithm is
proposed for the evaluation of the Renyi entanglement entropy(REE)
of quantum dimer model(QDM) at the Rokhsar-Kivelson(R-K) point. It
makes possible the evaluation of REE at the R-K point to the
thermodynamic limit for a general QDM. We apply the algorithm to QDM
on both triangular and square lattice as demonstrations and find the
REE on both lattices follow perfect linear scaling in the
thermodynamic limit, apart from an even-odd oscillation in the
latter case. We also evaluate the topological entanglement
entropy(TEE) on both lattices with a subtraction procedure. While
the expected TEE of $\ln2$ is clearly demonstrated for QDM on
triangular lattice, a strong oscillation of the result is found for
QDM on square lattice, which implies the relevance of boundary
perturbation in such a critical system.
\end{abstract}
\pacs{  75.10.-b, 73.43.-f, 71.27.+a}
 \maketitle

\section{I. Introduction}
Spin liquid phases in quantum magnet\cite{Frustrated} are exotic
state of matter that can realize novel quantum structures. The
topological order is one of such possibility. The topological order
manifests itself in both the nontrivial ground state degeneracy on
multiply connected manifold, and the topological contribution to the
entanglement entropy.

The quantum dimer model(QDM) is the simplest model system to
illustrate the spin liquid physics\cite{RK,Sondhi}. In a QDM, the
basic degree of freedom is the dimer living on the bonds of the
lattice. Each site of the lattice should be involved in one and only
one dimer. A dimer can be viewed as a spin singlet pair in the
resonating valence bond theory and can hop from bond to bond through
the kinetic term of QDM. At the so called the Rokhsar-Kivelson(R-K)
point, the ground state of QDM becomes the equal amplitude
superposition of all the allowed dimer coverings of the lattice.
Such a state can be viewed as a spin liquid if we interpret a dimer
as a spin singlet pair.

On multiply connected manifold, the Hilbert space of QDM factorizes
into distinct topological sectors that can not be connected by any
local Hamiltonian term to any finite order. As an example, for QDM
defined on a two dimensional torus, we can introduce two reference
lines around each of the two holes of the torus and classify a dimer
configuration into one of four sectors according to the parity of
the number of dimers that cross the two reference lines in this
configuration. It is clear that such parities are conserved by any
local Hamiltonian term and thus the Hilbert space factorizes into
four disconnected pieces. Since the different sectors are locally
indistinguishable, any local Hamiltonian defined on such factorized
Hilbert space should exhibit at least four-fold degeneracy. The
existence of such locally indistinguishable, but globally distinct
degenerate states on multiply connected manifold is one important
signature of a topological ordered state.

On bipartite lattice, additional subtlety emerges. Rather than the
parity of the crossing numbers, an integer-valued quantity is
conserved by all local Hamiltonian terms. As a result of such a
subtle difference, the QDM defined on frustrated and on unfrustrated
lattice have distinct properties. For example, while the QDM on the
triangular lattice has a short-ranged dimer correlation and exhibits
well defined $Z_{2}$ topological order at the R-K
point\cite{Sondhi,Furukawa}, the QDM at the R-K point on the square
lattice actually corresponds to a critical system with algebraic
dimer-dimer correlation and no topological order.

The topological entanglement entropy(TEE) is recently proposed as
another way to characterize a topological ordered
system\cite{Kitaev,Levin}. Unlike topological degeneracy, TEE can be
measured directly in the ground state wave function with no
reference to the excitation spectrum. For a gapped system, the
entanglement entropy generally satisfies the so called area law,
$S=\alpha A- \gamma$, in which $A$ is the area of the boundary
separating the subsystem from the environment. What makes a trivial
short range correlated state different from a topological ordered
state is that the latter exhibits a nonzero universal topological
contribution $\gamma$ to the entanglement entropy.

Besides being a useful way to characterize the topological order,
the information contained in the entanglement entropy can be used
more generally to diagnose the correlation in the system. In
critical system, correction to the linear scaling of entanglement
entropy is also generally
expected\cite{Pasquier,Alet,Oshikawa,Sandvik,Moore,Melko,Ju}. From
such corrections, important information about the low energy physics
of the system can be extracted.

For both of these purposes, an efficient algorithm is needed for the
evaluation of the entanglement entropy of QDM on large enough
subsystem. In the evaluation of entanglement entropy, the central
task is to enumerate the number of the dimer configurations in the
subsystem that is compatible with the fixed dimer configuration on
the boundary with the environment. For QDM defined on planar graph,
this can be done with a Fermioic representation of the dimer degree
of freedom\cite{Kasteleyn,Fisher,Stephan}. However, as the number of
dimer configurations on the boundary grows exponentially with the
size of the subsystem, such an algorithm is exponentially expensive
for large subsystem(see \cite{Stephan} for an exception in the
special geometry of an infinite cylinder). For non-planar lattice
such as three dimensional lattice, no efficient algorithm is known
to enumerate the dimer covering number.

For these reasons, the evaluation of TEE of QDM is limited to
subsystem of rather small size. As a result, the confirmation of the
$\ln2$ TEE for QDM on triangular lattice is still subjected to
rather large finite size uncertainty\cite{Furukawa}. The situation
become even worse for QDM on square lattice, for which the dimer
correlation is long-ranged and finite size effect is much stronger.
Furthermore, for QDM on 3D lattice and nonplanar 2D lattice, which
are of great theoretical interest, almost no result exist.

Thus, it is highly desirable to develop an efficient and easy to
implement algorithm to evaluate the entanglement entropy of a
general QDM at the R-K point. This is possible with the Monte Carlo
sampling, as the problem reduces to a classical statistical problem
at the R-K point. However, as the quantity to be averaged suffers
from an exponentially growing fluctuation with the growth of the
subsystem size, which is just the origin of the area law for
entanglement entropy, the convergence of the result becomes
exponentially slow.

In this paper, a re-weighting trick is proposed to overcome this
difficulty, which makes possible the evaluation of Renyi
entanglement entropy of general QDM at the R-K point to the
thermodynamic limit. We apply this algorithm to QDM on both the
square and the triangular lattice. We focus on two issues, namely
the scaling form of REE and the topological contribution TEE. We
find while the REE follows perfect linear scaling on triangular
lattice, a robust even-odd oscillation with the subsystem size in
REE is observed for QDM on square lattice. Nevertheless, we find the
REE for even and odd sized subsystem still follows perfect linear
scaling. We find the TEE on triangular lattice converge quickly to
the expected value of $\ln2$ with increasing subsystem size, while
that for square lattice exhibit strong oscillation and does not
converge.

This paper is organized as follows. In the next section, we describe
the way to evaluate the entanglement entropy with Monte Carlo
sampling in QDM. We also point out the origin of difficulty with the
conventional algorithm and introduce the re-weighting trick to solve
such difficulty. In section III, we demonstrate the new algorithm
with two examples, namely the QDM on square lattice and triangular
lattice, and present results on both the scaling form of REE and the
value of the topological contribution TEE calculated from a
subtraction procedure. In section IV, we present a discussion on the
result obtained and some outlook on the application of the new
algorithm to more general QDM.

\section{II. The Renyi entanglement entropy of QDM}
The ground state of a general QDM at the R-K point has the following
form
\begin{equation}
|\Psi\rangle=\sum_{C}|C\rangle.
\end{equation}
Here $|C\rangle$ denotes a general dimer covering of the lattice in
a specific topological sector. Each lattice site should be involved
in one and only one dimer in this covering.

To characterize the entanglement property of the system, we divide
it into two parts: a subsystem and the corresponding environment
that meet at their common boundary. The reduced density matrix of
the subsystem is then given by tracing out the degree of freedom in
the environment from the density matrix of the system
\begin{equation}
\rho_{s}=Tr_{e} | \Psi \rangle \langle \Psi |
\end{equation}
and the entanglement entropy of the Von Neumann type is defined as
\begin{equation}
S_{\mathrm{vN}}=-Tr\rho_{s}\ln\rho_{s}.
\end{equation}
However, it is in general quite hard to evaluate the Von Neumann
entropy numerically from this definition. A quantity that is much
easier to evaluate is the Renyi entropy. The Renyi entropy of order
$n$ is given by
\begin{equation}
S_{n}=-\frac{1}{n-1}\ln Tr\rho_{s}^{n}.
\end{equation}
The Von Neumann entropy is related to the Renyi entropy by
$S_{\mathrm{vN}}=\lim_{n\rightarrow 1}S_{n}$ and the two entropies
share many common properties in the long wave length limit. In
particular, it is found that the universal topological contribution
$\gamma$ extracted from $S_{n}$ is independent of the value of $n$
and agree with the value extracted from $S_{\rm{vN}}$\cite{Flammia}.

In the following, we will focus on the evaluation of $S_{2}$,
although the algorithm proposed below can be readily generalized to
evaluate $S_{n}$ with $n>2$. To evaluate $S_{2}$, we use the replica
trick\cite{Hastings,Zhang} and introduce an identical copy of the
system as its replica, which is divided in exactly the same manner
as the system into a subsystem and the corresponding environment.
The state vector of the whole system is then given by the direct
product of the state vector of the two copies,
$|\Psi\otimes\Psi\rangle$. It can be shown that
\begin{equation}
Tr\rho_{s}^{2}=\langle\Psi \otimes \Psi
|\widehat{\mathrm{SWAP}}|\Psi \otimes \Psi\rangle,
\end{equation}
in which $\widehat{\mathrm{SWAP}}$ denotes the operation of
exchanging all the degree of freedoms within the subsystem between
the system and its replica\cite{Hastings}. As the degree of freedom
of a QDM is defined on the bonds of the lattice, one should be
careful about the partition of the degree of freedoms. In Fig.1, a
partition on the square lattice is shown. In this partition, all
bonds whose midpoint that lie within the boundary line is defined to
be degree of freedom in the subsystem. Otherwise it is defined to be
degree of freedom of the environment.

\begin{figure}[h!]
\includegraphics[width=9cm,angle=0]{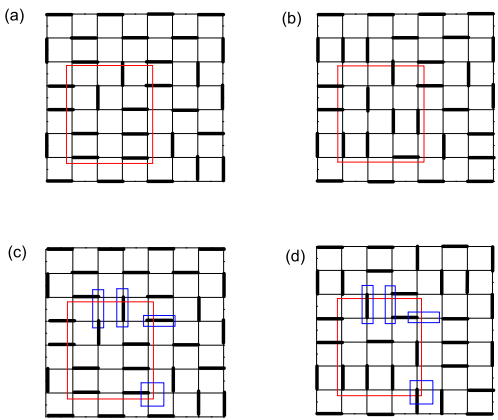}
\caption{Upper panel: an illustration of the partition of degree of
freedom and an example of swappable dimer configuration of the
system(a) and its replica(b). The subsystem is defined to be the
region within the red boundary lines. The dimers are shown with the
thick links. Lower panel: an example of non-swappable dimer
configuration of the system(c) and its replica(d). The locations of
local violation of the dimer constraint in the swapped configuration
are highlighted by the blue frames. In this example, $N_{err}=4$.}
\label{fig1}
\end{figure}

With such a partition, $Tr\rho_{s}^{2}$ can be represented in the
following form that is directly accessible to Monte Carlo sampling
\begin{equation}
Tr\rho_{s}^{2}=\frac{\sum_{C_{1},C_{2}}S(C_{1},C_{2})}{\sum_{C_{1},C_{2}}}.
\end{equation}
Here the summation is over all the dimer configurations in the
system ($C_{1}$) and its replica ($C_{2}$). $S(C_{1},C_{2})$ equals
to one if the swaped configuration still satisfy the dimer
constraint in both the system and its replica and is zero otherwise.
An example of swappable dimer configuration is shown in Fig.1a and
1b. A close inspection shows that $C_{1}$ and $C_{2}$ is swappable
only when the following two conditions are met. First, on each bond
in the subsystem that cross the boundary line, the dimer occupation
number in the system and its replica must be the same. Second, on
those bonds that cross the boundary line but belong to the
environment, the total number of dimers on those bonds that
terminating on the same site in the subsystem must be the same for
the system and its replica. Obviously, the number of such local
constraint grows linearly with the length of the boundary line. As a
result, the probability for two independent dimer configurations
$C_{1}$ and $C_{2}$ to be swappable decreases exponentially with the
length of the boundary. This is just the origin of the area law, and
it at the same time make clear the difficulty of evaluating the
entanglement entropy with Monte Carlo sampling, which amounts to
averaging a quantity with an exponentially large fluctuation.

However, the difficulty with the exponentially increasing
fluctuation in the sampling of entanglement entropy can be solved by
the following re-weighting trick. First, we re-write the sum in
Eq.(6) in the following form
\begin{equation}
Tr\rho_{s}^{2}=\frac{\sum_{C_{1},C_{2}}(r_{0})^{N_{err}}}{\sum_{C_{1},C_{2}}(r_{1})^{N_{err}}}.
\end{equation}
Here $r_{0}=0$ and $r_{1}=1$, $N_{err}$ is the number of times that
the local dimer constraint is violated in the swaped
configuration(See Fig.1c and 1d for a detailed illustration of the
meaning of $N_{err}$). It is then clear that the large fluctuation
in sampling $Tr\rho_{s}^{2}$ originates from the huge difference in
the distribution that appear in the denominator and the numerator. A
simple solution to this problem is to introduce a sequence of
intermediate values $r_{i}$ and rewrite the quantity in term of the
product of a serials of ratios,
\begin{equation}
Tr\rho_{s}^{2}=\prod_{i=1,m}\frac{\sum_{C_{1},C_{2}}(r_{i-1})^{N_{err}}}{\sum_{C_{1},C_{2}}(r_{i})^{N_{err}}},
\end{equation}
in which $r_{0}=0$, $r_{m}=1$ and $r_{i}<r_{j}$ for all $i<j$. If
the difference between the successive $r_{i}$ is chosen to be
sufficiently small, then each term in the product can be simulated
efficiently. In practice, we have divided the region $[0,1]$ into 10
equal intervals, which is found to be fine enough for good
convergence for subsystem with size up to $16\times16$, which is
already quite large. For even larger subsystem, one should introduce
more intervals. However, since the number of intervals needed only
scales linearly with the subsystem size, the thermodynamic limit can
be easily approached.

\begin{figure}[h!]
\includegraphics[width=8cm,angle=0]{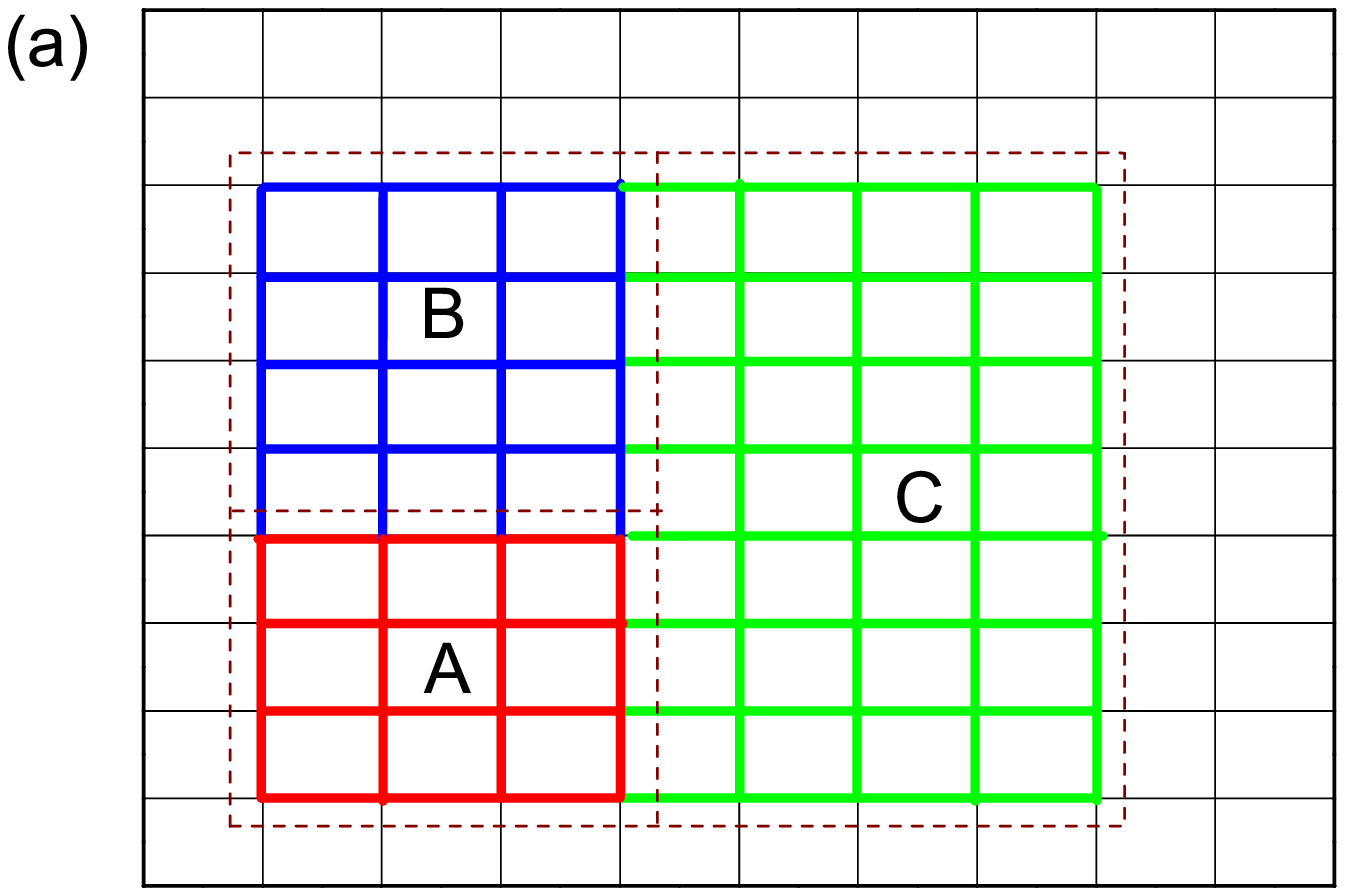}
\includegraphics[width=8cm,angle=0]{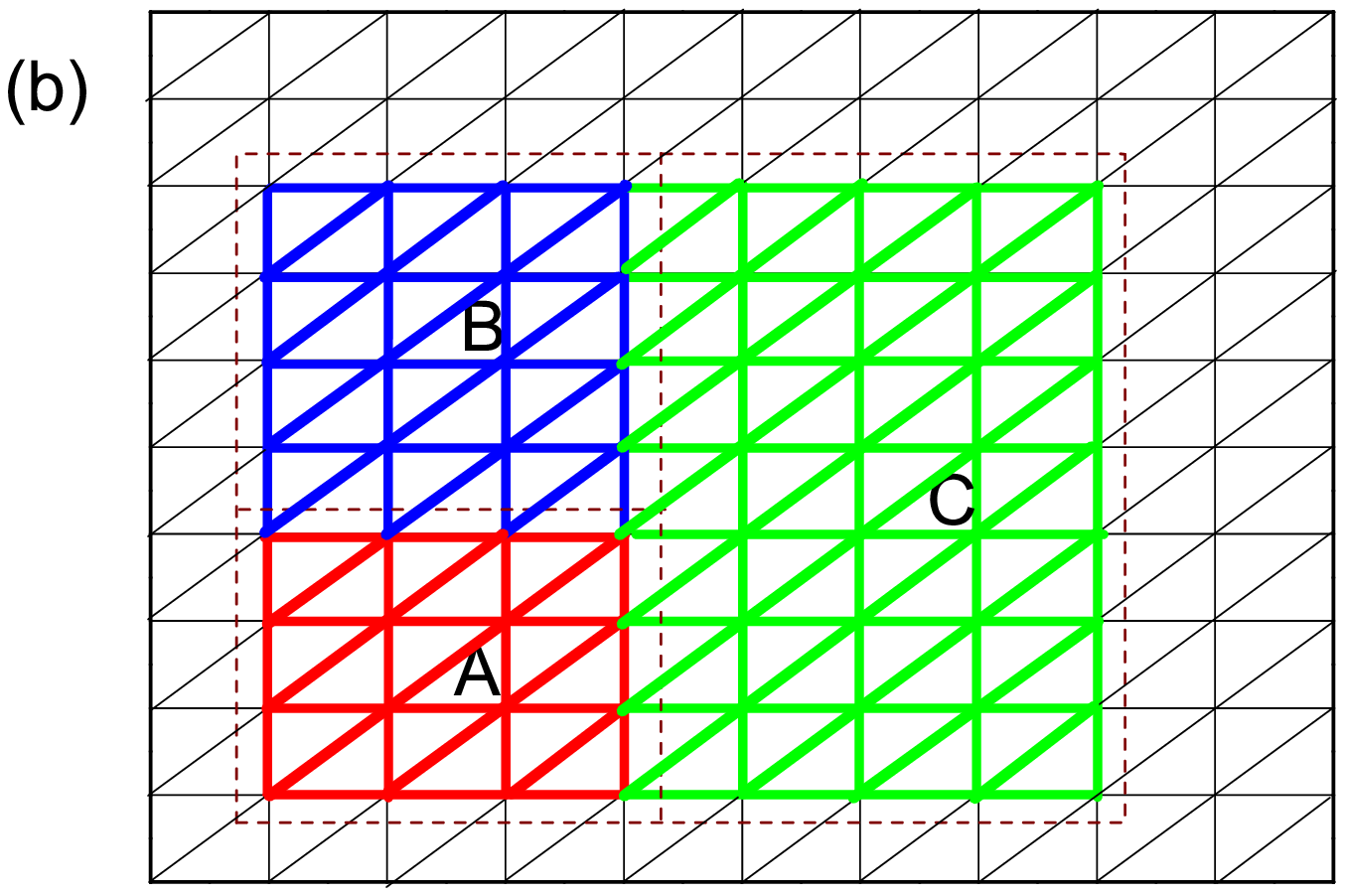}
\caption{The partition of the dimer degree of freedoms on the
lattice used to evaluate REE and TEE on (a) square lattice,
(b)triangular lattice. The subsystem used in the evaluation of REE
are all of A-type, for which all bonds that cross the boundary are
defined to be degree of freedom in the environment. The three
subsystems A, B and C separated by the dashed boundary lines will be
used later to extract the TEE. In this partition, a bond belongs to
a given subsystem when its midpoint belongs to that subsystem. Thus
all bonds in red belong to subsystem A, all bonds in blue belong to
subsystem B and all bonds in green belong to subsystem C. The
remaining bonds all belong to the environment. Note that although
subsystem A and B contains the same number of sites, they have
different number of bonds. } \label{fig2}
\end{figure}

\section{III. Application of the new algorithm}
\subsection{A. Scaling form of the Renyi entanglement entropy}

We apply the above algorithm to evaluate the Renyi entanglement
entropy of QDM at the R-K point on both triangular and square
lattice. The big system is defined on a torus with a fixed size of
$30\times30$ sites and the subsystem is chosen to contain
$n_{a}\times n_{a}$ lattice sites with $2\leq n_{a}\leq 16$. An
illustration of the subsystem is given in Fig.2. All bonds that
cross the boundary line are defined to be degree of freedom in the
environment in this calculation. We use $L=4n_{a}$ as a measure of
the boundary length. Note such a definition of boundary length has
an inherent ambiguity on a lattice.

The Renyi entanglement entropy $S_{2}$ of QDM on both the triangular
and the square lattice are shown in Fig.3. The error bar is smaller
than the symbol size. For the triangular lattice, a rather good
linear scaling of $S_{2}$ with $n_{a}$ is observed starting from a
rather small value of $n_{a}=2$, which is consistent with the short
ranged nature of the dimer correlation in the QDM on triangular
lattice. The Renyi entropy $S_{2}$ can thus be fitted well by the
following form
\begin{equation}
S_{2}=\alpha L-\beta.
\end{equation}
Note as a result of the ambiguity in the definition of the boundary
length, $\beta$ can not be directly interpreted as the topological
contribution to the entanglement entropy. The fitted values are
$\alpha=0.585$ and $\beta=2.85$.

For the square lattice, the result is markedly different. As can be
seen in Fig.3b, an even-odd oscillation in the Renyi entanglement
entropy is observed. Such an oscillation is also reported in other
recent studies of QDM and spinful short range RVB state on square
lattices and is attributed to the critical nature of the QDM on the
unfrustrated square lattice, which results in the relevance of the
boundary perturbation on $S_{n}$ for $n$ greater some critical
value\cite{Stephan}. For QDM on square lattice, it is found that
$n_{c}<2$. Thus $S_{2}$ is already susceptible to boundary
perturbation.

However, if we plot $S_{2}$ separately for subsystem with an even
and odd $n_{a}$, as is shown in Fig.4, it is clear that both results
follow perfect linear scaling for large $n_{a}$,
\begin{eqnarray}
S_{2}^{e}=\alpha_{e} L-\beta_{e}\\\nonumber
 S_{2}^{o}=\alpha_{o}L-\beta_{o},
\end{eqnarray}
where the offset $\beta_{e}$ and $\beta_{o}$ are found to be
negligible. It is also seen that the linear scaling starts at a much
larger value of $n_{a}=6$ than that on triangular lattice. This is
consistent with the fact that the QDM on square lattice has a
long-ranged dimer correlation.

\begin{figure}[h!]
\includegraphics[width=8cm,angle=0]{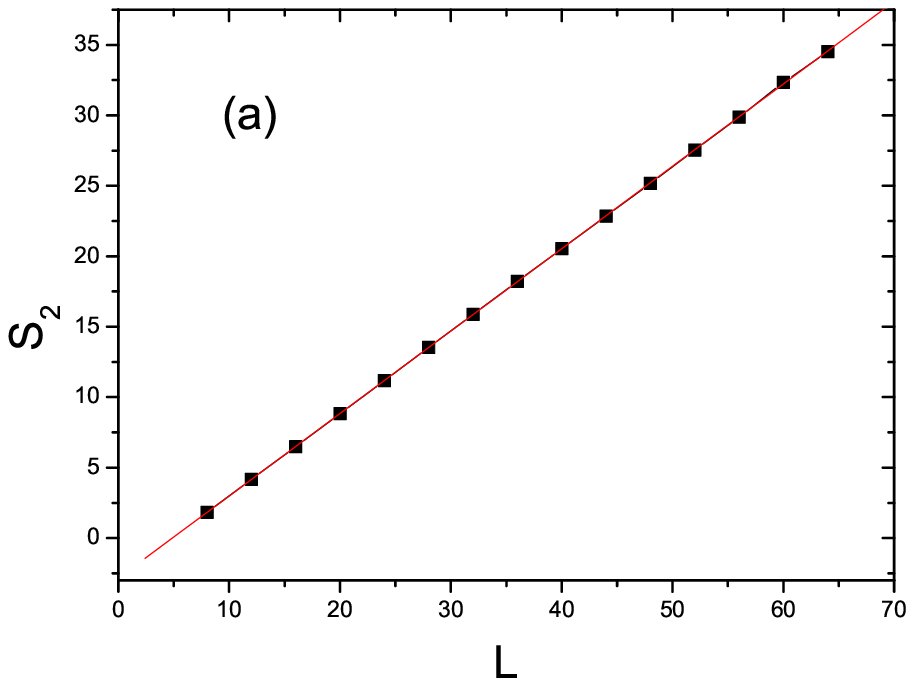}
\includegraphics[width=7cm,angle=0]{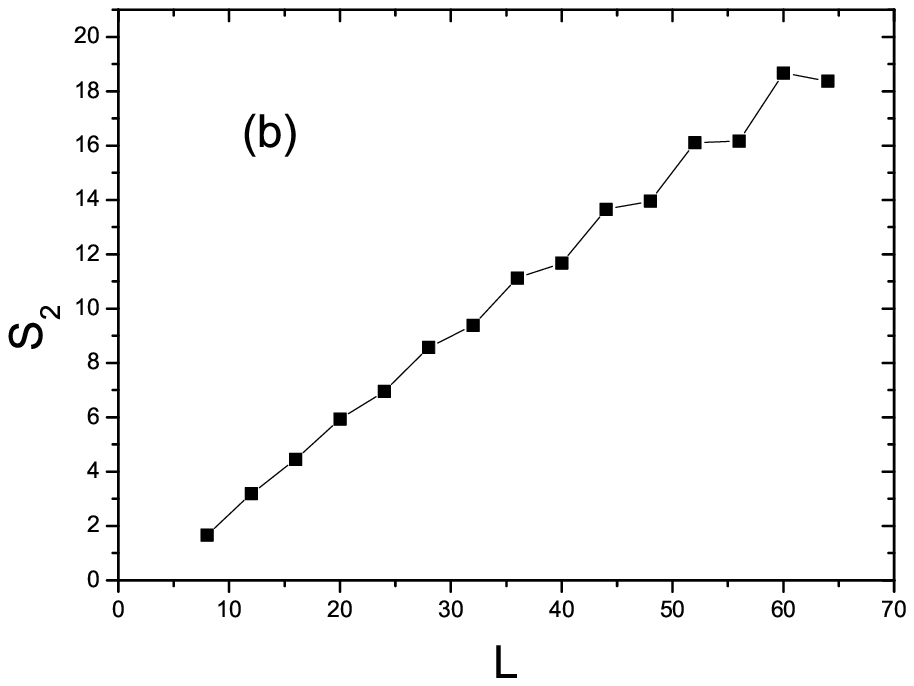}
\caption{The Renyi entanglement entropy $S_{2}$ of QDM as a function
of the boundary length $L=4n_{a}$. (a)triangular lattice, (b)square
lattice. The thin lines are guide to the eyes.} \label{fig2}
\end{figure}

\begin{figure}[h!]
\includegraphics[width=8cm,angle=0]{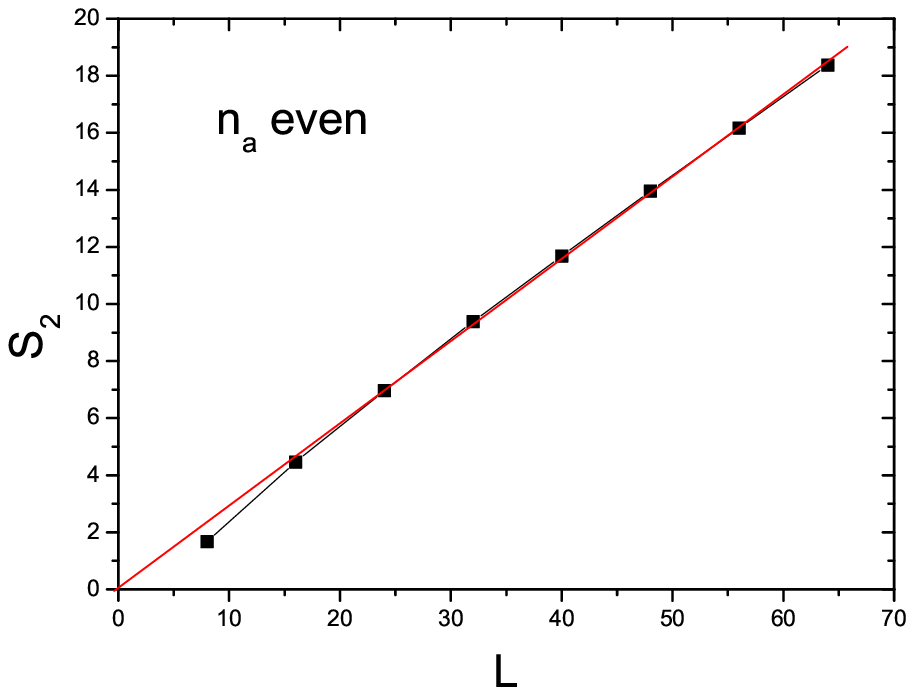}
\includegraphics[width=8cm,angle=0]{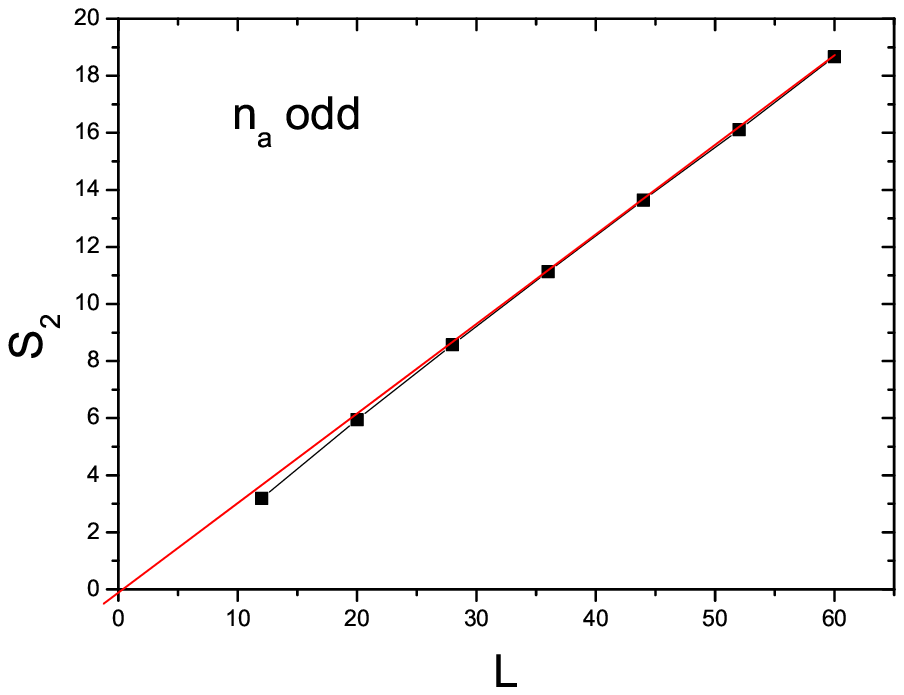}
\caption{The Renyi entanglement entropy $S_{2}$ of QDM on square
lattice as a function of the boundary length $L=4n_{a}$ for even(a)
and odd(b) $n_{a}$.  The thin lines are guide to the eyes.}
\label{fig3}
\end{figure}

\subsection{B. Topological entanglement entropy}

We now turn to the evaluation of TEE. For this purpose, we divide
the subsystem into three parts, A, B and C. Both A and B contain
$n_{a}\times n_{a}$ sites and C contains $n_{a}\times 2n_{a}$ sites.
Then the topological contribution to the entanglement entropy can be
extracted by subtracting out all the non-universal contributions
with the following combination
\begin{equation}
-\gamma=S_{2}^{A}+S_{2}^{B}+S_{2}^{C}-S_{2}^{AB}-S_{2}^{AC}-S_{2}^{BC}+S_{2}^{ABC}.
\end{equation}
Different from the spin system, care must be paid on the partition
of the dimer degree of freedom into the subsystems. For this
purpose, in Fig.2 we have plotted the bonds of different subsystems
with different colors. As a result of the difference in the
assignment of boundary degree of freedom, $S_{2}^{A}\neq S_{2}^{B}$
and $S_{2}^{AB}\neq S_{2}^{C}$. Thus, all the seven terms in Eq. 11
should now be calculated independently\cite{Zhang}.

\begin{figure}[h!]
\includegraphics[width=8cm,angle=0]{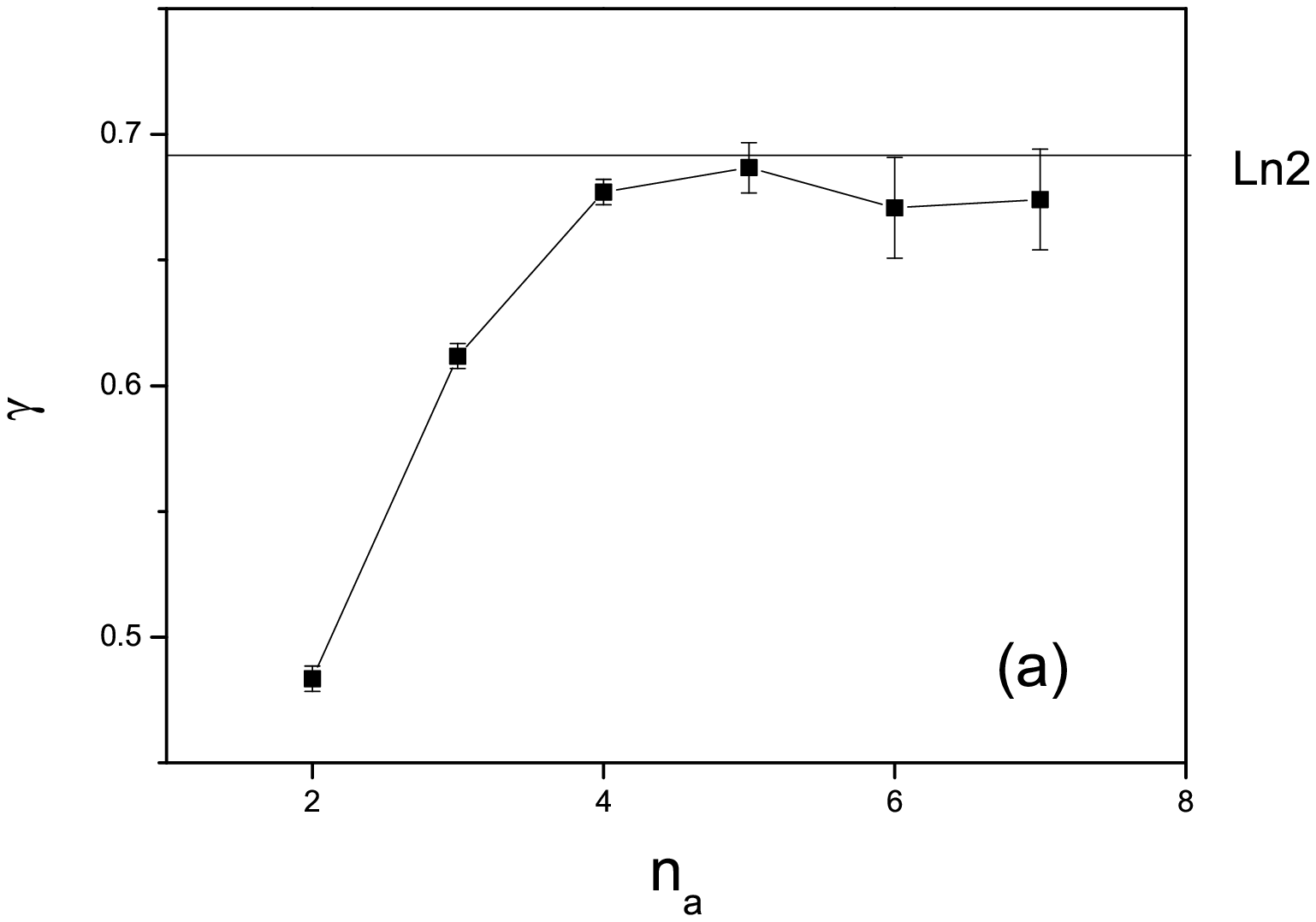}
\includegraphics[width=8cm,angle=0]{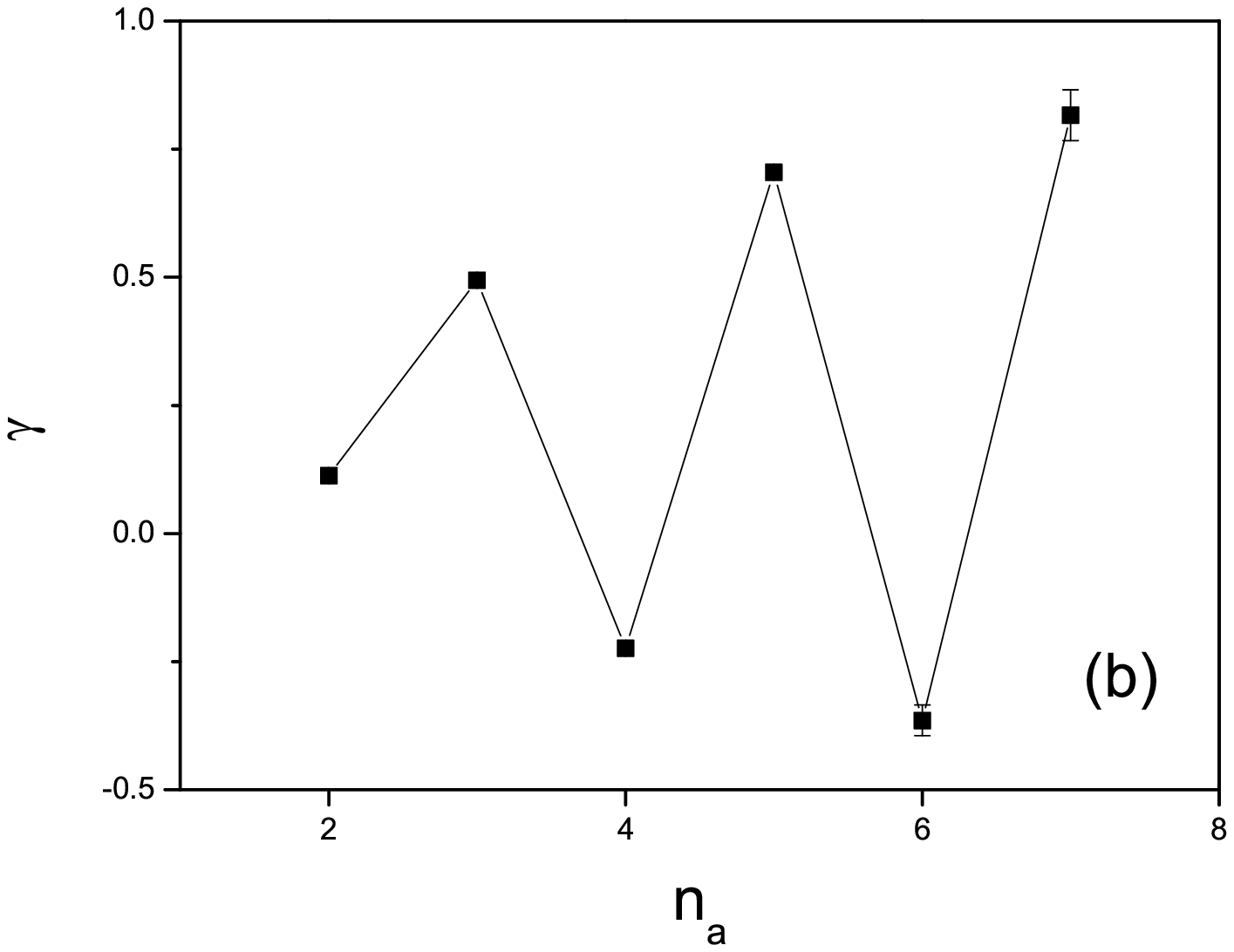}
\caption{The topological entanglement entropy extracted from Eq. 11
for QDM on triangular(a) and square(b) lattice. The thin lines are
gude to the eyes. } \label{fig5}
\end{figure}

We will first present the result of TEE for QDM on triangular
lattice, which is expected to show robust topological order and a
TEE of $\gamma=\ln 2$\cite{Sondhi,Furukawa}. This is confirmed by
our calculation. As is shown in Fig.5a, in which the TEE is plotted
as a function of $n_{a}$, $\gamma$ converges steadily to the
expected value of $\ln2$. Since the QDM on triangular lattice has a
rather short-ranged dimer correlation, we expect $\gamma$ to
converge to its thermodynamical limit at small $n_{a}$. From the
result it is also clear that the value extracted from linear
extrapolation of $S_{2}$($\beta=2.85$) can not be interpret as the
TEE and it is important to to perform the subtraction procedure as
indicated in Eq. 11 to remove the ambiguity in the definition of the
boundary length.

For the square lattice, the situation is quite different. As shown
in Fig.5b, the topological contribution extracted from Eq.11
exhibits strong oscillation with $n_{a}$ and never converge. This is
consistent with our understanding that the R-K point of the QDM on
square lattice describe a critical system, for which the topological
order is unstable. The strong oscillation is the result the even-odd
effect mentioned above, which implies again the relevance of the
boundary perturbation in such a critical system.

\section{IV. Discussions}

In this paper, we have presented a very efficient and simple to
implement Monte Carlo algorithm for the evaluation of the Renyi
entanglement entropy of QDM at the R-K point. The algorithm is
general and simple enough that it can be easily implemented for QDM
defined on a general lattice. A generalization of the algorithm for
$S_{n}$ with a higher order $n$ is also straightforward.

As a demonstration of the new algorithm, we have applied it to the
QDM defined on triangular and square lattice. On triangular lattice,
the $\ln 2$ TEE and the perfect linear scaling of REE in the
thermodynamic limit, which are expected from effective theory
arguments, are clearly demonstrated. On the other hand, an
interesting even-odd oscillation in REE is observed for QDM on
square lattice, which agrees with previous studies and implies the
relevance of boundary perturbation in such a critical system.
However, even if the dimer correlation is long ranged in such a
critical system, we find that the Renyi entropy still exhibit a
linear scaling with boundary length in the thermodynamic limit if we
consider separately the Renyi entropy of even and odd sized
subsystems. No logarithmic correction to the scaling is found.

The power of this new algorithm is of course not limited to the
above demonstrations. As QDM now becomes an important playground for
the construction and detection of novel quantum state of
matter\cite{Frustrated}, it is interesting to apply the new
algorithm to more general QDM systems. In this respect, it is
particularly interesting to study QDM defined on three dimensional
lattice and to see how the ideas for 2D systems can be generalized
to 3D system and what kind of new structure can emerge. As an
concrete example, QDM defined on the pyrochlore lattice is
especially attractive since it is closely related to the Kagome
lattice quantum spin liquid on the one hand\cite{Yan} and to the
spin ice and monopole excitation on the other hand\cite{Morris}.

After the completion of this work, we notice a recent work on the
evaluation of the Renyi entanglement entropy of QDM at the R-K
point\cite{Whaley}. In their work, a ratio estimator trick first
proposed in the study of spinful RVB state\cite{Hastings} is
adopted. This trick, working like peeling the onion, involves the
calculation of the ratios of REE on a serials of subsystems with
decreasing sizes. Our algorithm, on the other hand, only involves
the evaluation of the REE at a fixed subsystem size and is thus much
easier to implement for more complex models.

This work is supported by NSFC Grant No. 10774187 and National Basic
Research Program of China No. 2007CB925001 and No. 2010CB923004. Tao
Li acknowledges the hospitality received from SISSA, Trieste, where
part of this work is done.

Note: Qiang Han is now an undergraduate student of the department
and should not to be mistaken as the professor in the department
with the same name.

\end{document}